\documentclass[reprint,
superscriptaddress,
amsmath,amssymb,
aps,nofootinbib,
prd,onecolumn
]{revtex4-2}

\usepackage{graphicx}
\usepackage[usenames,dvipsnames]{xcolor}
\definecolor{bluecite}{HTML}{0875b7}
\usepackage[colorlinks=true,urlcolor=Emerald,linkcolor=bluecite,citecolor=bluecite]{hyperref}
\usepackage{acronym}
\usepackage{MnSymbol}
\usepackage{physics} 
\usepackage{float}
\usepackage{dsfont} 
\usepackage{mathrsfs} 

\newcommand{\sgn}{\mathrm{sign}} 

\newcommand{\msf}[1]{\mathsf{#1}} 
\newcommand{\ii}{\mathrm{i}} 
\newcommand{\ttt}[1]{\texttt{#1}} 

\begin{document}
	
	\title{Investigating the Araki-Uhlmann relative entropy between two coherent states in relativistic Quantum Field Theory}

	\author{Jo{\~a}o Gabriel Carib{\'e}}
	\email{joaogcaribe@cbpf.br}
	\affiliation{CBPF - Centro Brasileiro de Pesquisas Físicas, Rua Dr. Xavier Sigaud 150, 22290-180, Rio de Janeiro, Brazil}

	\author{Marcelo S.  Guimaraes,}
	\email{msguimaraes@uerj.br}
	\affiliation{UERJ - Universidade do Estado do Rio de Janeiro,	Instituto de Física - Departamento de Física Teórica - Rua São Francisco Xavier 524, 20550-013, Maracanã, Rio de Janeiro, Brazil}
	
	\author{Itzhak Roditi}
	\email{roditi@cbpf.br}
	\affiliation{CBPF - Centro Brasileiro de Pesquisas Físicas, Rua Dr. Xavier Sigaud 150, 22290-180, Rio de Janeiro, Brazil}
	
	\author{Silvio P. Sorella}
	\email{silvio.sorella@fis.uerj.br}
	\affiliation{UERJ - Universidade do Estado do Rio de Janeiro,	Instituto de Física - Departamento de Física Teórica - Rua São Francisco Xavier 524, 20550-013, Maracanã, Rio de Janeiro, Brazil}
	
	
	\date{\today}

	\begin{abstract}
		A numerical setup for investigating the Araki-Uhlmann relative entropy between two coherent states is presented for a scalar massive Quantum Field Theory in ($1+1$)-dimensional Minkowski spacetime. These states are constructed using smeared Weyl operators compactly supported in two diamond regions belonging to the right Rindler wedge. Using this setup, we verified the known properties of the relative entropy, namely: positivity, increase with the size of the spacetime regions considered, decrease with the increase of the mass parameter. A linear increase with respect to the spatial distance between the two diamond regions is also observed. 
	\end{abstract}
		\maketitle
	

\section{Introduction}\label{sec:intro}
	
The Araki-Uhlmann relative entropy~\cite{Araki:1975zw,Araki:1976zv,Uhlmann:1976me} quantifies the distinguishability between two given quantum states and can be seen as a generalization of the standard quantum relative entropy. It plays an important role in the current understanding of entanglement in Quantum Field Theory (QFT) and has been used in a number of applications ranging from cosmology to information theory~\cite{Hiai:1991mxv,Witten:2018zxz,Casini:2022rlv,Hollands:2019czd,DAngelo:2021yat,Ciolli:2021otw,Casini:2022rlv,Galanda:2023vjk,Garbarz:2022wxn,Floerchinger:2020ogh,Dowling:2020nxc}.
\\\\In a recent work, a numerical setup to compute the relative entropy for a scalar massive QFT~\cite{Guimaraes:2025cqt} was devised to study its behavior with respect to the field mass parameter as well as verify its independence from the choice of the test functions used to smear out the quantum field.  Moreover, that setup also provided an opportunity to explicitly confirm the positivity of the relative entropy as well as its monotonic increase with the size of the spacetime regions under consideration. All these issues have been scrutinized by looking at the relative Araki-Uhlmann entropy between the vacuum state and a coherent state. 
\\\\In the present work, we extend the analysis done in~\cite{Guimaraes:2025cqt} to the case of two coherent states localized in diamond shaped regions on the right Rindler wedge. In addition to recovering all properties mentioned in the previous paragraph, we also have found that the Araki-Uhlmann entropy increases linearly with respect to the spatial distance between the two diamonds. To the best of our knowledge, this is the first quantitative investigation of the properties of the relative entropy between two coherent states.
\\\\This work is organized as follows. In Sect.\eqref{sec.background} we present a minimal construction of a massive scalar QFT using the usual canonical quantization procedure and construct its coherent states using Weyl operators. Sect.\eqref{sec:Araki-Uhlmann relative entropy} summarizes the construction of the Araki-Uhlman relative entropy for two coherent states. Sect.\eqref{num} is devoted to the presentation of the numerical procedure and contains our results. Finally, we present our conclusions in Sect.\eqref{conc}. 
			
\section{The coherent states of a free quantum scalar field}\label{sec.background}

\subsection{The smeared quantum scalar field and its canonical commutation relations}
The line element of a flat (1+1)-dimensional Minkowski spacetime $\mathfrak{S}$ in Minkowski coordinates $\left\{t\in\mathbb{R}\,,x\in\mathbb{R}\right\}$ is given by
\begin{equation}\label{eq: Line element}
    \dd s^2 = \dd t^2 - \dd x^2.
\end{equation}
On $\mathfrak{S}$ we consider a massive free quantum scalar field $\hat{\varphi}$ obeying the Klein-Gordon equation $\Box\hat{\varphi} = 0$. The general real-valued solution for that equation can be written as
\begin{equation} \label{eq: scalar field}
\hat{\varphi}(\msf{x}) = \int_{\mathbb{R}} \frac{\dd k}{2 \pi} \frac{1}{2 \omega_{k}} \left( e^{-\ii\msf{k}_\mu \msf{x}^\mu} \hat{a}_k + e^{\ii \msf{k}_\mu \msf{x}^\mu} \hat{a}^{\dagger}_k \right), 
\end{equation}
where $\msf{x} = (t,x)$ is a spacetime point, $\msf{k} = (\omega_{k}, k)$ is the wave vector obeying the dispersion relation $\omega_{k} = \sqrt{k^2 + m^2}$ which arises from the Klein-Gordon equation, and $\msf{k}_\mu \msf{x}^\mu = \omega_{k} t - k x$, where we used the Einstein summation convention, assumed from now on. The creation and annihilation operators $\hat{a}^{\dagger}_k$ and $\hat{a}_k$, respectively, obey the usual canonical commutation relations\footnote{The normalization we use for the creation and annihilation operators differs from the one in~\cite{Peskin2018} by a factor of $\sqrt{2\omega_{k}}$. For this reason, the commutator in Eq.\eqref{eq: Canonical commutator 1} picks an extra factor of $2\omega_{k}$ when compared to the corresponding expression in that reference.}:
\begin{equation}\label{eq: Canonical commutator 1}
    [\hat{a}_k, \hat{a}^{\dagger}_{k'}] = 2\pi \, 2\omega_{k} \, \delta(k - k')\mathds{1}
\end{equation}
and
\begin{equation}\label{eq: Canonical commutator 2}
    [\hat{a}_k, \hat{a}_k] = [\hat{a}^{\dagger}_k, \hat{a}^{\dagger}_{k'}] = 0,
\end{equation}
where $\mathds{1}$ is the identity operator and $\delta$ is the Dirac-delta distribution.
\\\\
We remark that $\hat{\varphi}(\msf{x})$ is an operator-valued distribution~\cite{Haag:1992hx} which should be smeared against a suitable test function to properly define the corresponding field operator. For that end we consider a test function $f$ from the Schwartz space $\mathcal{S}(\mathbb{R}^2)$ to define the smeared field operator as
\begin{equation}
    \hat{\varphi}[f]\equiv \int_{\mathfrak{S}} \dd^2\msf{x} \hat{\varphi}(\msf{x})f(\msf{x})\,,\,f\in \mathcal{S}(\mathbb{R}^2).
\end{equation}
The corresponding smeared creation operator is defined as
\begin{equation}
    \hat{a}^{\dagger}_{f} \equiv \int_{\mathbb{R}}\frac{\dd k}{2\pi}\frac{1}{2\omega_{k}}\tilde{f}(\msf{k})\hat{a}^{\dagger}_k,
\end{equation}
where
\begin{equation}
    \tilde{f}(\msf{k}) = \int_{\mathfrak{S}}\dd^2\msf{x}e^{\ii \msf{k}_\mu \msf{x}^{\mu}}f(\msf{x})
\end{equation}
is the Fourier transform of $f$. Similarly, the annihilation operator corresponding to $\hat{\varphi}[f]$ is defined as
\begin{equation}
    \hat{a}_{f} \equiv \int_{\mathbb{R}}\frac{\dd k}{2\pi}\frac{1}{2\omega_{k}}\tilde{f}^{*}(\msf{k})\hat{a}_k,
\end{equation}
where the asterisk denotes complex conjugation\footnote{While $f$ is a real-valued function, its Fourier transform $\tilde{f}$ is, in general, complex-valued.}.
In $\hat{a}^{\dagger}_{f}$ and $\hat{a}_{f}$, the smeared field operator can be written as
\begin{eqnarray}
    \hat{\varphi}[f] = \hat{a}_{f} + \hat{a}^{\dagger}_{f}
\end{eqnarray}
and the canonical commutation relations take the forms
\begin{equation}\label{eq: Smeared canonical commutator 1}
    [\hat{a}_f, \hat{a}^{\dagger}_{g}] = \braket{f}{g} = \int_{\mathbb{R}}\frac{\dd k}{2\pi}\frac{1}{2\omega_{k}} \tilde{f}^*(\msf{k})\tilde{g}(\msf{k})\mathds{1}
\end{equation}
and
\begin{equation}\label{eq: Smeared canonical commutator 2}
    [\hat{a}_f, \hat{a}_g] = [\hat{a}^{\dagger}_f, \hat{a}^{\dagger}_{g}] = 0,
\end{equation}
where $g\in \mathcal{S}(\mathbb{R}^2)$ and $\braket{f}{g}$ is the Lorentz-invariant inner product between the test functions $f$ and $g$. In the configuration space, this inner product can be written as
\begin{equation} 
    \braket{f}{g} = \frac{i}{2} \Delta_{PJ}(f,g) + H(f,g) \;, \label{csp}
\end{equation} 
where $\Delta_{PJ}(f,g)$ and $H(f,g)$ are the smeared Pauli-Jordan and Hadamard distributions presented in App.\eqref{appA}.
\\\\
The vacuum state of $\hat{\varphi}$ is unique and is defined by
\begin{equation}
    \hat{a}_{f}\ket{\Omega} = 0\, ,\forall\,f\in \mathcal{S}(\mathbb{R}^2).
\end{equation}

With these elements, we proceed to the construction of the coherent states of $\hat{\varphi}$.

\subsection{Coherent states}\label{sec:Coherent states}
We begin by defining the von Neumann algebra\footnote{See \cite{Guimaraes:2024mmp} for a concise introduction to the subject. } ${\cal M}(O)$ generated by the smeared unitary Weyl operators 
\begin{equation}
{\hat{\cal W}}_f \equiv e^{i \hat{\varphi}[f]} \;, \qquad {\hat{\cal W}}_f \;{\hat{\cal W}}_f^\dagger = {\hat{\cal W}}_f^{\dagger}  \;{\hat{\cal W}}_f = \mathds{1} \;, \label{Wf}
\end{equation} 
where $f\in\mathcal{S}(\mathbb{R}^2)$ is a test function supported in an open region $O$ of $\mathfrak{S}$. Hence,
\begin{equation} 
{\cal M}(O) = \left\{\; {\hat{\cal W}}_f,  \; supp(f) \subseteq O \; \right\}^{"}
\end{equation}
where the symbol ${"}$ stands for the bi-commutant.
\\\\

It follows from the Reeh-Schlieder Theorem \cite{Haag:1992hx} that the vacuum state $\ket{\Omega}$ of $\hat\varphi$ is both cyclic and separating for the von Neumann algebra ${\cal M}(O)$. The corresponding coherent state $\ket{f}$ is obtained by acting with the Weyl operators ${\hat{\cal W}}_f$ on the vacuum state:
\begin{equation}
|f\rangle = {\hat{\cal W}}_f \ket{\Omega} = e^{i \hat{\varphi}[f] } \ket{\Omega}  \;. \label{cfst}
\end{equation}
Due to the following property of the Weyl operators:
\begin{equation} 
{\hat{\cal W}}_f {\hat{\cal W}}_{f'} = e^{-\frac{i}{2} \Delta_{PJ}(f,f')} {\hat{\cal W}}_{(f+f')} \;, \label{walg} 
\end{equation} 
coherent states are, as the vacuum state $\ket{\Omega}$, both cyclic and separating for ${\cal M}(O)$.

\section{The Araki-Uhlmann relative entropy for two coherent states}\label{sec:Araki-Uhlmann relative entropy}
Consider two coherent states $|f\rangle$ and $|g\rangle$ constructed from the test functions $f\in\mathcal{S}(\mathbb{R}^2)$ and $g\in\mathcal{S}(\mathbb{R}^2)$, respectively. The Araki-Uhlmann relative entropy  \cite{Araki:1975zw,Araki:1976zv,Uhlmann:1976me}  between these two states is given by 
\begin{equation} 
S(f\vert g) = - \bra{f}\log\hat{\Delta}_{fg}\ket{f}, \label{AU}
\end{equation}
where $\hat{\Delta}_{fg}$ is the relative Tomita-Takesaki modular operator \cite{tomita1967canonical,Takesaki:1970aki,Witten:2018zxz}, obtained from the anti-linear unbounded operator $\hat{s}_{fg}$, defined by the closure of the map 
\begin{equation} 
\hat{s}_{fg}\; \hat{\omega}\; |f\rangle = \hat{\omega}^{\dagger} \;|g\rangle \;, \qquad \forall \hat{\omega} \in {\cal M} \;. \label{stt}
\end{equation}
The operator $\hat{\Delta}_{fg}$ is thus obtained from the polar decomposition of $\hat{s}_{fg}$, namely,
\begin{equation} 
\hat{s}_{fg} = \hat{J}_{fg} \; \hat{\Delta}^{1/2}_{fg} \;, \label{pdd}
\end{equation}
where the operator $\hat{J}_{fg}$ is anti-unitary, known as the modular conjugation. The modular operator $\hat{\Delta}^{1/2}_{fg}$ is self-adjoint and positive. \\\\Before going further, it is worth to rederive an important property of $\hat{\Delta}_{fg}$. For that, let $\hat{s}_\Omega$ be the Tomita-Takesaki anti-linear operator corresponding to the vacuum state $\ket{\Omega}$, {\it i.e.}  
\begin{equation} 
\hat{s}_{\Omega} \; \hat{a} \ket{\Omega} = \hat{a}^{\dagger} \ket{\Omega} \;, \qquad \hat{s}^2_\Omega =1 \;, \qquad \hat{s}_\Omega = \hat{J}_\Omega \; \Delta^{1/2} _{\Omega} \;. \label{sO}
\end{equation}
From 
\begin{equation} 
\hat{s}_{fg} \; \hat{a} \ket{\Omega} = \hat{s}_{fg}\; \hat{a} \;{\hat{\cal W}}_f^\dagger {\hat{\cal W}}_f \ket{\Omega} = \hat{s}_{fg} \; \hat{a} \; {\hat{\cal W}}_f^{\dagger} |f\rangle = {\hat{\cal W}}_f \; \hat{a}^\dagger \; |g\rangle = {\hat{\cal W}}_f \; \hat{a}^{\dagger} \; {\hat{\cal W}}_{g} \; \ket{\Omega} = {\hat{\cal W}}_f \; \hat{s}_{\Omega} \; {\hat{\cal W}}^{\dagger}_g \; \hat{a} \; \ket{\Omega} \;, \label{ffr}
\end{equation}
it follows the important relation 
\begin{equation} 
\hat{s}_{fg} = {\hat{\cal W}}_f \; \hat{s}_{\Omega} \; {\hat{\cal W}}^{\dagger}_g \;, \label{fff1}
\end{equation}
from which 
\begin{equation} 
\hat{\Delta}_{fg} = \hat{s}^\dagger_{fg} \hat{s}_{fg} = {\hat{\cal W}}_{g} \;\hat{s}^\dagger_\Omega \;{\hat{\cal W}}^\dagger_f \;{\hat{\cal W}}_f \;\hat{s}_\Omega \;{\hat{\cal W}}^\dagger_g = {\hat{\cal W}}_g \;\hat{\Delta}_\Omega \; {\hat{\cal W}}_g^\dagger \;. \label{dd1}
\end{equation}
Moreover, following \cite{Longo:2019mhx}, an operational definition of the relative entropy is achieved through the expression, 
\begin{equation} 
S(f\vert g) = i \frac{d\;}{d s}  \langle f|\;\hat{\Delta}^{is}_{fg} \; |f\rangle\Big|_{s=0}  \;, \label{AUL}
\end{equation}
with 
\begin{equation} 
\hat{\Delta}^{is}_{fg} = e^{is \log(\hat{\Delta}_{fg})} \;, \qquad s \in {\mathbb{R}} \;. \label{expis}
\end{equation}
Thus, from \eqref{dd1}, one has 
\begin{eqnarray} 
{\cal S}(f|g) & =  & i \frac{d\;}{d s}  \langle f|\;{\hat{\cal W}}_g \;\hat{\Delta}^{is}_{\Omega} \;{\hat{\cal W}}_g^\dagger \; |f\rangle\Big|_{s=0}  =  i \frac{d\;}{d s}  \langle \Omega |\;{\hat{\cal W}}_f^\dagger \; {\hat{\cal W}}_g \;\hat{\Delta}^{is}_{\Omega} \;{\hat{\cal W}}_g^\dagger \; {\hat{\cal W}}_f \ket{\Omega}\Big|_{s=0} \nonumber \\
\; &=& i \frac{d\;}{d s}  \langle \Omega |\;{\hat{\cal W}}_f^\dagger \; {\hat{\cal W}}_g \; \hat{\Delta}^{is}_{\Omega} \;{\hat{\cal W}}_g^\dagger \; (\hat{\Delta}^{is}_{\Omega})^\dagger \;\hat{\Delta}^{is}_{\Omega}\; {\hat{\cal W}}_f \; (\hat{\Delta}^{is}_{\Omega})^\dagger \;\ket{\Omega}\Big|_{s=0} 
\end{eqnarray} 
where use has been made of the modular invariance of the vacuum state, 
\begin{equation} 
\hat{\Delta}_\Omega^{is} \; \ket{\Omega} = \ket{\Omega}  \;. \label{inOm}
\end{equation}
Furthermore, from the Bisognano-Wichmann results \cite{Bisognano:1975ih}, the action of the modular flow $\hat{\Delta}_{\Omega}^{is}$ on the Weyl operators is known and given by 
	\begin{equation} 
		\hat{\Delta}_{\Omega}^{is}\; {\hat{\cal W}}_f \; (\hat{\Delta}_{\Omega}^{is})^\dagger = e^{i \hat\varphi[\delta^{is}f]} \;, \qquad \delta^{is}f(\msf{x}) = f(\Lambda_{-s}\, \msf{x}) \;, \label{da}
	\end{equation}
	where $\Lambda_s$ stands for a Lorentz boost: 
	\begin{align}
		\Lambda_s\msf{x}: 
		\left\{
		\begin {aligned}
		&  \;\; x'  =  \cosh(2\pi s) \;x - \sinh(2 \pi s) \; t, \\
		&  \;\;  t'  =  \cosh(2\pi s) \;t - \sinh(2 \pi s) \; x.          
	\end{aligned}
	\right. \label{bst}
\end{align}
Collecting everything and making use of 
\begin{equation} 
\langle \Omega|\; e ^{i \hat{\varphi}[f]}\; e^{i \hat{\varphi}[g]} \;\ket{\Omega} = e^{-\frac{i}{2} \Delta_{PJ}(f,g)} \; e^{-\frac{1}{2} ||f+g||^2} \;, \label{exp}
\end{equation}
where 
$||\cdot||^2$ is the norm induced by the Lorentz invariant inner product from Eq.\eqref{eq: Smeared canonical commutator 1}, one arrives at the final equation 
\begin{equation}
S(f\vert g) = - \frac{1}{2} \Delta_{PJ}(f-g, f'_s - g'_s) \Big|_{s=0} \;, \qquad f'_s = \frac{df}{ds}, \label{ffA}
\end{equation}
where $f_s(\msf{x})\equiv f(\Lambda_{-s}\msf{x})$. This equation, will be the starting point of the numerical study which will be presented in the next section.
%
%
%
%
%
%
%
%
%
%
%
\section{Numerical setup and results} \label{num}
\begin{figure}
\centering
    \includegraphics[width=.85\textwidth]{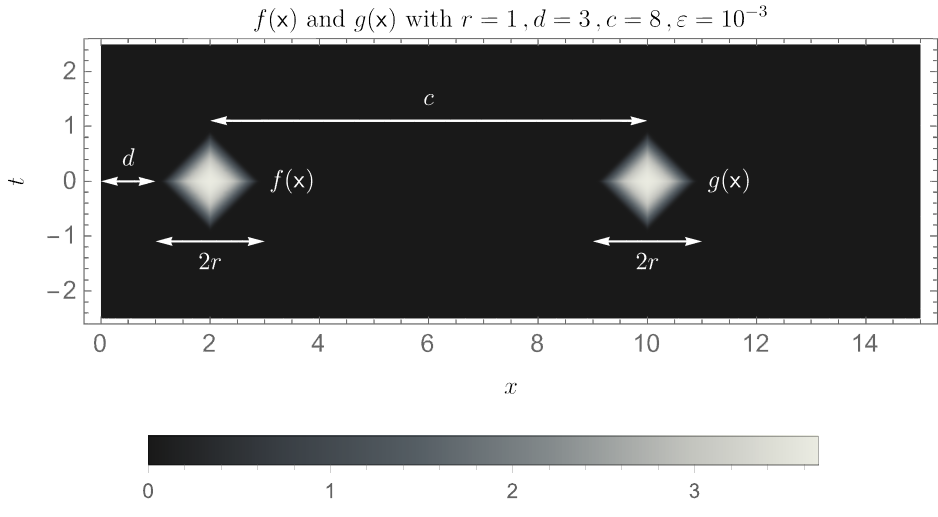}
    \caption{ The test functions $f(\msf{x})$ and $g(\msf{x})$ (Eqs.\eqref{ft} and~\eqref{gt}) are plotted as a function of $\msf{x} = (t,x)$. Here we used $r = 1$, $d = 3$, $c = 8$ and $\varepsilon = 10^{-3}$. Note the diamond-shape of the regions where $f$ and $g$ have support.
    }
    \label{fg}
\end{figure}

Here, we consider two coherent states $\ket{f}$ and $\ket{g}$ constructed as in Sect.\ref{sec:Coherent states} with test functions $f$ and $g$ compactly supported on diamond-shaped regions $A,B\subset\mathfrak{S}$ located in the right Rindler wedge $R = \left\{\msf{x} = (t,x) \in\mathfrak{S}\, ,\,x\geq \abs{t}\right\}$ of the (1+1)-dimensional Minkowski spacetime $\mathfrak{S}$. 
\\\\
Region $A$ is chosen as the casual diamond centered at $\mathsf{x}_{A} = (t = 0, x = d + r)$, where $d$ is the $x$-coordinate distance between the origin $\mathsf{x}_0 = (t = 0, x = 0)$ and the left-vertex of $A$ and $2r$ is the distance between its left- and right-vertices. Explicitly, $A = \left\{\msf{x} = (t,x)\in\mathfrak{S}\, ,\, \abs{d+r-x} + \abs{t} \leq r\right\}$. We then choose the test function $f$ as
\begin{align}
f(\msf{x})=
\left\{
\begin{aligned}
& {\rm exp}\bigg[{\frac{-1}{(\sqrt{r^2+\varepsilon^2}+\varepsilon)^2-\lambda(\msf{x},\varepsilon)^2}}\bigg] \;, \;\;\;\;\;
& \lambda_A(\msf{x},\varepsilon)\leq \sqrt{r^2+\varepsilon^2}+\varepsilon,\;\\
&0,\;\;\;\; {\rm elsewhere},
\end{aligned}
\right.\label{ft}
\end{align}
where $\lambda_A(\msf{x},\varepsilon) \equiv \sqrt{(d+r-x)^2+\varepsilon^2}+\sqrt{t^2+\varepsilon^2}$ is an auxiliary function and $\varepsilon$ is a smoothing parameter that interpolates the support of $f$ between region $A$ for $\varepsilon = 0$ and a circle with radius $r$ centered at $\msf{x}_A$ for $\varepsilon\to\infty$.
\\\\
Region $B$ is chosen as the causal diamond with the same shape as that of $A$ but centered on $\mathsf{x}_{B} = (t = 0, x = d + r + c)$, i.e., at a $x$-coordinate distance $c$ away from the center of $A$. Explicitly, $B = \left\{\msf{x} = (t,x)\in\mathfrak{S}\, ,\, \abs{c+d+r-x} + \abs{t} \leq r\right\}$. The test function $g$ is then chosen as
\begin{align}
g(\msf{x})=
\left\{
\begin{aligned}
& {\rm exp}\bigg[{\frac{-1}{(\sqrt{r^2+\varepsilon^2}+\varepsilon)^2-\lambda_B(\msf{x},\varepsilon)^2}}\bigg] \;, \;\;\;\;\;
& \lambda_B(\msf{x},\varepsilon)\leq \sqrt{r^2+\varepsilon^2}+\varepsilon,\;\\
&0,\;\;\;\; {\rm elsewhere},
\end{aligned}
\right. \label{gt}
\end{align} 
where $\lambda_B(\msf{x},\varepsilon) \equiv \sqrt{(c+d+r-x)^2+\varepsilon^2}+\sqrt{t^2+\varepsilon^2}$ is an auxiliary function. An illustration of the test functions $f$ and $g$ is presented in Fig.\ref{fg} where the roles of the parameters $d$, $r$ and $c$ are also depicted. For both $f$ and $g$, the smoothing parameter $\varepsilon$ must be larger than zero to ensure that $f\in \mathcal{S}(\mathbb{R}^2)$.
\\\\
With the test functions defined, they can be used in the expression for relative entropy $S(f\vert g)$ between $\ket{f}$ and $\ket{g}$ from Eq.~\eqref{ffA}. The integrals to be evaluated are $\Delta_{PJ}(f-g,f'_s-g'_s)|_{s=0}$. Due to the difficulties in deriving a closed analytical expression for the Fourier transform of the compact supported test functions $f$ and $g$, these integrals have been directly evaluated in the configuration space (see Eq.~\eqref{mint}). In all cases, we used the \ttt{QuasiMontecarlo} method implemented in Mathematica~\cite{Mathematica} using \ttt{Maxpoints}$\;= 10^5$ when studying the $m$- and $r$-dependence of $S(f\vert g)$ and \ttt{Maxpoints}$\;= 10^8$ when analyzing its $c$-dependence.

\subsection{The dependence on the field mass \texorpdfstring{$m$}{m}}

\begin{figure}[t!]
\centering
\includegraphics[width=.75\textwidth]{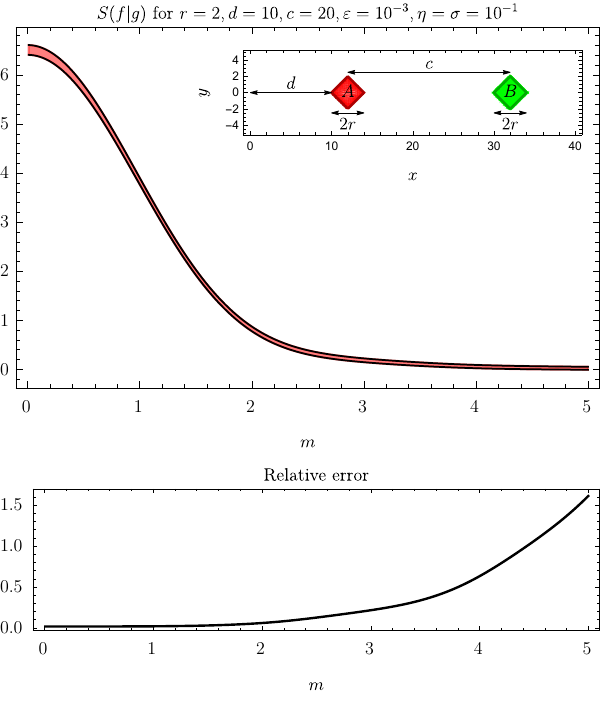}
    \caption{
    The top plot shows the Araki-Uhlman relative entropy $S(f\vert g)$ (Eq.\eqref{ffA}) between the coherent states $\ket{f}$ and $\ket{g}$ (Sect.\ref{sec:Coherent states}) constructed from the test functions $f(\msf{x})$ (Eq.\eqref{ft}) and $g(\msf{x})$ (Eq.\eqref{gt}) as a function of the mass $m$ of the scalar field $\hat{\varphi}$.
    The inset shows the causal diamonds $A$ and $B$, where the test functions $f$ and $g$ are, respectively, supported. The bottom plot shows the relative errors from the numerical evaluation of $S(f\vert g)$.
    The upper and lower black lines around the 
    shaded region represent the error bands within which $S(f\vert g)$ lies. Being a quantification of the distinguishability between $\ket{f}$ and $\ket{g}$, the Araki-Uhlman relative entropy in the top-plot shows that these states become monotonically less distinguishable as the mass $m$ of $\hat\varphi$ increases.
    }
\label{mass}
\end{figure}

The free scalar field theory has a single parameter: the field mass $m$ which plays a role in the relative entropy $S(f\vert g)$ between two coherent coherent states $\ket{f}$ and $\ket{g}$. As can be seen in Eq.\eqref{ffA}, it is determined by the smeared Pauli-Jordan distribution. In the limits of low- and large-mass, one can use the asymptotic limits of the Bessel function of the first kind $J_0$~\cite[Chap 10.2]{NIST:DLMF}, respectively given by\footnote{These asymptotic limits hold as long as $\sqrt{t^2-x^2}$ remains finite, which is ensured by the fact that the test functions $f$ and $g$ are compactly supported within a causal diamond of finite vertex-to-vertex length $2r$.}
\begin{equation}
    J_0(\msf{x}) = 1 + \mathcal{O}(m^2),\, m\in\mathbb{R}_{>0}, \, m\to 0
\end{equation}
and
\begin{equation}
    J_0(\msf{x}) = \frac{(1+i)}{2\sqrt{m\pi} (t^2-x^2)^{3/4}} \left(e^{-i m \sqrt{t^2-x^2}}\sqrt{t^2-x^2}-\ii\sqrt{t^2-x^2} e^{i m \sqrt{t^2-x^2}}\right) + \mathcal{O}(m^{-1}),\, m\in\mathbb{R}_{>0}, \, m\to \infty,
\end{equation}
where $\msf{x} = (t,x) \in \mathfrak{S}$. We complement these limits with the remark that on the light-cone ($x^2 - t^2 = 0$), $J_0(\msf{x}) = 1$ for any $m\in\mathbb{R}_{>0}$. 
\\\\
Therefore the large-mass limit of the smeared Pauli-Jordan distribution (Eq.\eqref{mint}) reads
\begin{equation}
    \Delta_{PJ}(f,g) = 0 + \mathcal{O}(m^{-1/2}),\, m\in\mathbb{R}_{>0}, \, m\to \infty,
\end{equation}
which straightforwardly implies that $S(f\vert g) \to 0$ in this limit. On the other hand, in the low-mass limit,
\begin{equation}
    \Delta_{PJ}(f,g) =  -\frac{1}{2}\int_{\mathfrak{S}} d^2\msf{x} d^2\msf{y}\, \sgn(t-t')\theta\left((t-t')^2 - (x-x')^2\right) f(\msf{x})g(\msf{y}) + \mathcal{O}(m^2),\, m\in\mathbb{R}_{>0}, \, m\to 0,
\end{equation}
where $\msf{x} = (t,x)$ and $\msf{y}=(t',x')$. Hence, in this limit, the Araki-Uhlman relative entropy $S(f\vert g)$ is finite, since $f,g\in\mathcal{S}(\mathbb{R}^2)$. 
\\\\
To probe the intermediary mass regime, which interpolates between the low- and large-mass limits, we numerically performed the integrals that result from substituting the test functions $f$ and $g$ from Eqs.\eqref{ft} and~\eqref{gt}, respectively, into the expression for $S(f\vert g)$ from Eq.\eqref{ffA}. Here we fixed the parameters $r = 2$, $d = 10$, $c = 20$, $\varepsilon = 10^{-3}$ and $\eta = \sigma = 10^{-1}$ and varied $m$ from $10^{-2}$ to $5$ in steps of $10^{-2}$. The result is presented in Fig.\eqref{mass} together with an estimate of the relative error in the numerical integration. This figure shows that the Araki-Ulhman relative entropy decreases monotonically as the mass parameter increases, interpolating between the low-mass limit, where $S(f\vert g)  = 6.5$ and the large-mass limit, where it is zero. 
\\\\
That $S(f\vert g)$ tends to zero in the large-mass limit can be traced to the fact that as the mass increases, the support of the Pauli-Jordan distribution dwindles to a region around the boundary of the light-cone. On top of that, in contrast to the (3+1)-dimensional Minwkoski case, in (1+1) dimensions the Pauli-Jordan is not supported on the light-cone. Therefore, it becomes zero everywhere in the infinite-mass limit, which implies that $S(f\vert g)\to 0$ as well. Hence, in that limit $\ket{f}$ and $\ket{g}$ are indistinguishable.
\\\\
We can understand such indistinguishability by analyzing the smeared Hadamard distribution (Eq.\eqref{mint}), which encodes the state-dependence. In the infinite-mass limit it goes to zero, implying that the state-dependence is washed out. Consequently, a given state $\ket{f}$ becomes indistinguishable from any other state $\ket{g}$.
\\\\
Since these explanations do not depend on the form of $f$ and $g$ but only on the fact that they are elements of $S(\mathbb{R}^2)$, the monotonically decreasing behavior of $S(f\vert g)$ with the mass is also independent of the choice of $r$,$d$ and $c$.
\\\\
It is important to remark that while in the low-mass limit $S(f\vert g)$ remains finite because the Pauli-Jordan distribution is finite in this limit, the Hadamard distribution diverges as $\sim\ln(m)$. Therefore, $m$ can be as small as desired, as long as it remains larger than zero. Indeed, as can be seen in Fig.\eqref{mass}, this is the case where $S(f\vert g)$ is maximum, which implies that $\ket{f}$ and $\ket{g}$ are as distinguishable as possible for the given test-functions $f$ and $g$.

\subsection{The dependence on the properties of the coherent states}

\begin{figure}[ht!]
\centering
\includegraphics[width=.75\textwidth]{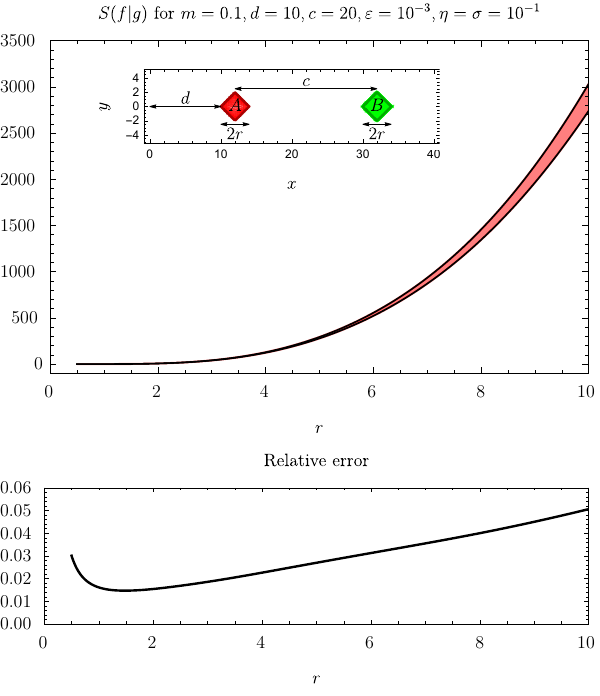}
    \caption{
    The top plot shows the Araki-Uhlman relative entropy $S(f\vert g)$ (Eq.\eqref{ffA}) between the coherent states $\ket{f}$ and $\ket{g}$ (Sect.\ref{sec:Coherent states}) constructed from the test functions $f(\msf{x})$ (Eq.\eqref{ft}) and $g(\msf{x})$ (Eq.\eqref{gt}) as a function of $r$, the parameter that controls the size of the diamond-shaped regions $A$ and $B$ where each test function is supported.
    The bottom plot shows the relative errors from the numerical evaluation of $S(f\vert g)$ as function of $r$.
    The upper and lower black lines around the 
    shaded region represent the error bands within which $S(f\vert g)$ lies. 
    Here we see that $\ket{f}$ and $\ket{g}$ becomes increasingly distinguishable as the diamond size $r$ increases.
    }
\label{size}
\end{figure}
\begin{figure}[ht!]
\centering
\includegraphics[width=.75\textwidth]{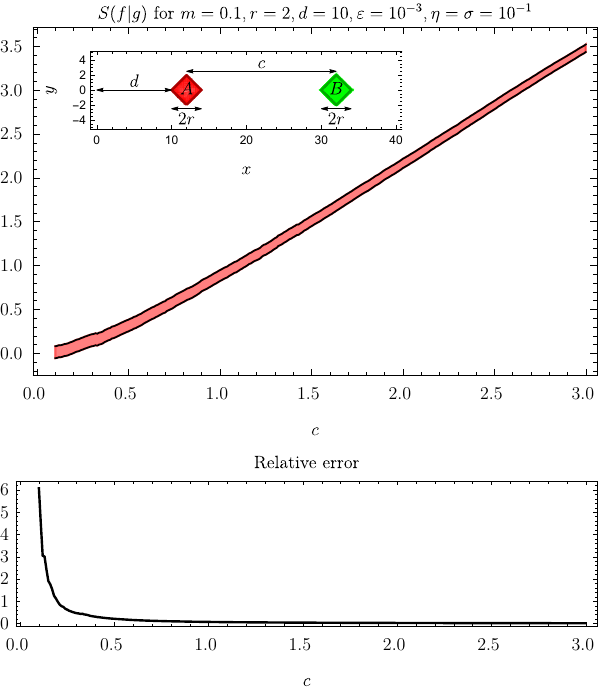}
    \caption{
    The top plot shows the Araki-Uhlman relative entropy $S(f\vert g)$ (Eq.\eqref{ffA}) between the coherent states $\ket{f}$ and $\ket{g}$ (Sect.\ref{sec:Coherent states}) constructed from the test functions $f(\msf{x})$ (Eq.\eqref{ft}) and $g(\msf{x})$ (Eq.\eqref{gt}) as a function of the coordinate separation $c$ between the centers of the diamond-shaped regions $A$ and $B$ where $f$ and $g$ are supported, respectively.
    The bottom plot shows the relative errors from the numerical evaluation of $S(f\vert g)$ as function of $c$. The large relative error as $c\to 0$ is expected since in that limit $S(f\vert g)\to 0$ as well and such value is difficult to obtain from a numerical integration performed with the Quasi-Monte Carlo method.
    The upper and lower black lines around the 
    shaded region represent the error bands within which $S(f\vert g)$ lies. 
    Here we see that $\ket{f}$ and $\ket{g}$ becomes increasingly distinguishable as the diamond separation $c$ increases, with a transition for a super-linear regime for $c \ll r$ to a linear regime when $c \gtrsim r$.
    }
\label{distance}
\end{figure}

The state-dependence of the relative entropy $S(f\vert g)$ (see Eq.\eqref{ffA}) between two coherent quantum states $\ket{f}$ and $\ket{g}$ of the massive quantum scalar field $\hat{\varphi}$ is implied from the test-functions $f(\msf{x})$ and $g(\msf{x})$ used to define them. Here we use the compactly-supported test-functions from Eqs.\eqref{ft} and~\eqref{gt}.
\\\\
To study the $r$-dependence of $S(f\vert g)$, we fixed $m = 10^{-1}$, $d = 10$, $c = 20$, $\varepsilon = 10^{-3}$ and $\eta=\sigma=10^{-1}$ and varied $r$ from $0.5$ to $10$ in steps of $10^{-2}$. The result is presented in Fig.\eqref{size} and shows that the relative entropy increases monotonically with $r$, meaning that $\ket{f}$ and $\ket{g}$ become increasingly distinguishable. This is a well-known property of the relative entropy, reviewed in~\cite{Witten:2018zxz} and explicitly verified here for the case of two coherent states.
\\\\
Finally, to analyze the $c$-dependence of $S(f\vert g)$, we fixed $m = 10^{-1}$, $r = 2$, $d = 10$, $\varepsilon = 10^{-3}$ and $\eta=\sigma=10^{-1}$ and varied $c$ from $0.1$ to $3$ in steps of $10^{-2}$. The result is presented in Fig.\eqref{distance}. Here we see that the relative entropy increases monotonically with the separation between the diamond-shaped regions. It has two distinct regimes: For $c \ll r$, the increase in $S(f\vert g)$ is faster than linear. When $c \gtrsim r$, it transitions into a linear regime.
\\\\
This behavior can be understood as follows: When $c = 0$ both test-functions coincide, i.e., $f = g$. As a consequence, $\ket{f}=\ket{g}$ which trivially implies that these states are indistinguishable. This explains why $S(f\vert g)\to 0$ when $c\to 0$, as seen in Fig.\eqref{distance}. As $c$ increases the test-functions become increasingly distinct. Consequently, $\ket{f}$ and $\ket{g}$ become increasingly distinguishable which is reflected in the increase in $S(f\vert g)$ Fig.\eqref{distance}. In short, $\ket{f}$ and $\ket{g}$ become mo distinguishable as the spatial distance between them increases. To our knowledge, this is the first explicit result on this interesting feature of the relative entropy.

\section{Conclusion}\label{conc}

In this work, we extended the numerical analysis of the Araki-Uhlmann relative entropy started in \cite{Guimaraes:2025cqt} to the case of two coherent states $\ket{f}$ and $\ket{g}$, localized in diamond regions on the right Rindler wedge; see Fig.\eqref{fg}.
Besides recovering the expected behaviors with respect to the mass parameter, Fig.\eqref{mass}, and to the size of the diamond regions, Fig.\eqref{size}, a linearly increasing behavior with respect to the spatial distance $c$ between the two diamonds has been detected, Fig.\eqref{distance}. 
To our knowledge, this is the first quantitative result on this interesting feature of the Araki-Uhlmann relative entropy, meaning that the more the spatial distance $c$ increases the more the two coherent states become distinguishable.

\section*{Acknowledgments}
The authors would like to thank the Brazilian agencies Conselho Nacional de Desenvolvimento Científico e Tecnológico (CNPq), Coordenação de
Aperfeiçoamento de Pessoal de Nível Superior - Brasil (CAPES) and Fundação Carlos Chagas Filho de Amparo à Pesquisa do Estado do Rio de Janeiro (FAPERJ) for financial support. In particular, S. P.~Sorella, I.~Roditi, and M. S.~Guimaraes are CNPq researchers under contracts 301030/2019-7, 311876/2021-8, and 309793/2023-8, respectively.


\appendix

\section{Explicit expressions of the Pauli-Jordan and Hadamard distributions}\label{appA}

The smeared Pauli-Jordan and Hadamard distributions are defined as
\begin{align}
\Delta_{PJ}(f,g) &=  \int \! d^2\msf{x} d^2\msf{y} f(\msf{x}) \Delta_{PJ}(\msf{x}-\msf{y}) g(\msf{y}) \;,  \nonumber \\
H(f,g) &=  \int \! d^2\msf{x} d^2\msf{y} f(\msf{x}) H(\msf{x}-\msf{y}) g(\msf{y})\;, \label{mint}
\end{align}
where $f,g\in\mathcal{S}(\mathbb{R}^2)$ are test functions. Explicitly, the Pauli-Jordan $\Delta_{PJ}(\msf{x}-\msf{y})$ and Hadamard $H(\msf{x}-\msf{y})$ distributions are given by
\begin{eqnarray} 
\Delta_{PJ}(\msf{x}) & =&  -\frac{1}{2}\;{\rm sign}(t) \; \theta \left( \lambda(t,x) \right) \;J_0 \left(m\sqrt{\lambda(t,x)}\right) \;, \nonumber \\
H(\msf{x}) & = & -\frac{1}{2}\; \theta \left(\lambda(t,x) \right )\; Y_0 \left(m\sqrt{\lambda(t,x)}\right)+ \frac{1}{\pi}\;  \theta \left(-\lambda(t,x) \right)\; K_0\left(m\sqrt{-\lambda(t,x)}\right) \;, \label{PJH}
\end{eqnarray}
where 
\begin{equation} 
\lambda(t,x) = t^2-x^2 \;, \label{ltx}
\end{equation}
$J_0$ and $Y_0$ are the Bessel functions of the first kind~\cite[Sec 10.2]{NIST:DLMF}, $K_0$ is the modified Bessel function of the second kind~\cite[Sec 10.25]{NIST:DLMF} and $m$ is the field's mass parameter.




\bibliography{refs}
%

\end{document}